\documentclass[pra,twocolumn,showpacs]{revtex4}
\usepackage{amsmath}
\usepackage{amssymb}
\usepackage{epsfig}

\begin{document}

\title{\bf Non-stationary vortex ring in a Bose-Einstein condensate with Gaussian density}
\author{Victor P. Ruban}
\email{ruban@itp.ac.ru}
\affiliation{L.D. Landau Institute for Theoretical Physics RAS, Moscow, Russia} 

\date{\today}

\begin{abstract}
The local induction equation, approximately describing dynamics of a quantized vortex filament in a 
trapped Bose-Einstein condensate in the Thomas-Fermi regime on a spatially nonuniform density background 
$\rho({\bf r})$ and taking dimensionless form 
${\mathbf R}_t=\varkappa {\mathbf b}+[\nabla\ln\rho({\mathbf R})\times {\boldsymbol \tau}]$ 
(where $\varkappa$ is a local curvature of the filament, ${\mathbf b}$ is the unit binormal vector, 
and ${\boldsymbol \tau}$ is the unit tangent vector), is shown to admit 
a finite-dimensional reduction if the density profile is an isotropic Gaussian, $\rho\propto\exp(-|{\bf r}|^2/2)$. 
The reduction corresponds to a geometrically perfect vortex ring centered at position ${\bf A}(t)$, 
with orientation and size both determined by a vector ${\bf B}(t)$. Parameters 
${\bf A}$ and ${\bf B}$ exhibit the same dynamics as velocity and position of a Newtonian particle do in 3D:
$\dot {\bf A}={\bf B}/|{\bf B}|^2-{\bf B}$, and $\dot {\bf B}={\bf A}$.
\end{abstract}
\maketitle

{\bf Introduction}.
Dynamics of quantum vortices in a trapped atomic Bose-Einstein condensate with spatially inhomogeneous
equilibrium density  $\rho({\mathbf r})$ is an important and interesting problem for physical
experiment as well as for the theory (see
\cite{F2009,SF2000,FS2001,R2001,AR2001,GP2001,A2002,RBD2002,AD2003,AD2004,rings-2004,SR2004,D2005,
Kelvin_vaves,ring_istability,v-2015,reconn-2017,top-2017}, and many references therein). 
In general case it is impossible to separate potential and vortical excitations of 
the condensate, but if the condensate at zero temperature is in the Thomas-Fermi regime, 
then one can use the ``anelastic'' hydrodynamic approximation to study vortex motion theoretically
\cite{SF2000,FS2001,R2001,A2002,SR2004,R2017-1,R2017-2}. The approximation works well when vortex core width 
$\xi= \xi_0[\rho({\mathbf r})/\rho_0]^{-1/2}$ is much smaller than a typical scale of 
inhomogeneity and vortex size $R_*$, while the maximum of vortex line curvature 
$\varkappa_{\rm max}$ is of order $R_*$. If, besides that, configuration of a single vortex line 
is far from self-intersections, then a simple mathematical model is applicable, the local induction equation
\cite{SF2000,FS2001,R2001}
\begin{equation}
{\mathbf R}_t\big|_{\rm normal}=\frac{\Gamma \Lambda}{4\pi}\Big(\varkappa {\mathbf b}
+[\nabla\ln\rho({\mathbf R})\times {\boldsymbol \tau}]\Big),
\label{LIA}
\end{equation}
where ${\mathbf R}(\beta,t)=(X(\beta,t), Y(\beta,t), Z(\beta,t))$ is a geometric shape of the filament 
depending on arbitrary longitudinal parameter
$\beta$ and time $t$, coefficient $\Gamma=2\pi\hbar/m$ is the velocity circulation quantum for atomic mass $m$,
$\Lambda=\ln(R_*/\xi)\approx$ const is a large logarithm, $\varkappa$ is a local curvature of the vortex line, 
${\mathbf b}$ is the unit binormal vector, and ${\boldsymbol \tau}$ is the unit tangent vector.

To make formulas clean, below we use dimensionless quantities, so that ${\Gamma \Lambda}/{4\pi}=1$, $R_*\sim 1$.
It is a well known fact that in the case $\rho =$ const, the local induction equation is reduced by the Hasimoto
transform \cite{Hasimoto} to  one-dimensional (1D) focusing nonlinear Schr\"odinger equation, so the vortex line 
dynamics against a uniform background is nearly integrable. For nonuniform densities investigation of this model 
is still in the very beginning, but some interesting results have already been  obtained 
\cite{Kelvin_vaves,ring_istability,R2016-1,R2016-2,R2017-3}. In particular, exact solutions in the form of a moving
straight vortex for anisotropic Gaussian density profiles were studied in Ref.\cite{R2016-2}.
Very recently, parametric instabilities of vortex ring on a $z$-periodic density background  
were revealed for definite ring sizes, while in a harmonically trapped condensate, parametric instabilities 
take place at definite values of trap anisotropy \cite{R2017-3}.

In this brief note, some new exact solutions of Eq.(\ref{LIA}) will be discussed, for central symmetric 
Gaussian density $\rho\propto\exp(-|{\bf r}|^2/2)$. In this case the anelastic theory works
inside domain of a radius $R_{\max}$ satisfying condition $[\xi_0/R_{\max}]\exp(R_{\max}^2/4)\lesssim 1$.
The equation takes form
\begin{equation}
{\mathbf R}_t=\varkappa {\mathbf b}-[{\mathbf R}\times {\boldsymbol \tau}]
\label{LIA_Gaussian}
\end{equation}
and admits non-stationary solutions corresponding to motion and rotation of geometrically perfect ring. 

But before going to the main subject of the work, we would like to say some more words about the local 
induction model (\ref{LIA}) in the context of Bose-Einstein condensates.

{\bf New derivation of Eq.(\ref{LIA})}.
Two different, mutually independent approaches were used in \cite{SF2000,FS2001,R2001} to derive Eq.(\ref{LIA}), 
but there still exists the third, more direct way to obtain it. Indeed, the basic Gross-Pitaevskii equation 
for wave function $\Psi({\bf r},t)$ of a dilute gas Bose-Einstein condensate is of the canonical form
\begin{equation}
i\hbar\Psi_t={\delta{\cal H}}/{\delta\Psi^*},
\label{Psi_variat}
\end{equation}
with the Hamiltonian given by the well-known Gross-Pitaevskii energy functional,
\begin{equation}
{\cal H}=\int \Big[\frac{\hbar^2}{2m}|\nabla\Psi|^2
+[V({\bf r})-\mu]|\Psi|^2+\frac{g}{2}|\Psi|^4\Big]d^3{\bf r}.
\end{equation}

In the ``anelastic'' hydrodynamic approximation,  function $\Psi$ is determined exclusively 
by vortex line configuration, so existence of a functional $\Psi({\bf r}, \{{\bf R}(\beta)\})$ is implied. 
For $R_*\gg\xi$ and $1/\varkappa_{\rm max}\gg \xi$, and closely to vortex line, $\Psi$ is approximately two-dimensional,
so that in the locally perpendicular plane ${\bf r}_\perp =(r_\perp\cos\varphi, r_\perp\sin\varphi)$ we have
\begin{equation}
\Psi\approx \Psi_v({\bf r}_\perp)=F_{[\rho]}(r_\perp)e^{i\varphi},
\end{equation}
where $\Psi_v$ corresponds to a straight vortex on a uniform background, with a local value of vortex-free density
$\rho=\rho({\bf R})=m|\Psi_0({\bf R})|^2\approx m[\mu-V({\bf R})]/g$. 

Due to Eq.(\ref{Psi_variat}), the following relation takes place,
\begin{equation}
i\hbar\int\Big[\Psi_t\frac{\delta\Psi^*}{\delta {\bf R(\beta)}}
-\Psi^*_t\frac{\delta\Psi_v}{\delta {\bf R(\beta)}}\Big]d^3{\bf r}=
\frac{\delta{\cal H}}{\delta{\bf R(\beta)}}.
\label{hydro}
\end{equation}
One can estimate the left hand side of this equation similarly to Appendix B of Ref.\cite{BN2015}, 
using just basic local properties of functional $\Psi({\bf r}, \{{\bf R}(\beta)\})$, since the integral 
is mainly contributed by close vicinity of the filament where 
\begin{eqnarray}
\Psi_t&\approx&-{\bf R}_t\cdot\nabla\Psi_v({\bf r}_\perp),\\
\frac{\delta\Psi^*}{\delta{\bf R}}&\approx&-\delta(r_\parallel-R_\parallel)|{\bf R}_\beta|\nabla\Psi^*_v({\bf r}_\perp).
\end{eqnarray}
Now we substitute these expressions into Eq.(\ref{hydro}), use formula for a double vector cross-product, 
and integrate over the angular coordinate $\varphi$. Thus we arrive at
\begin{equation}
2\pi\hbar|{\bf R}_\beta|[{\boldsymbol \tau}\times{\bf R}_t]\int_0^\infty \frac{d F_{[\rho]}^2}{d r_\perp} d r_\perp
\approx\frac{\delta{\cal H}}{\delta{\bf R(\beta)}}.
\end{equation}
An essential point here is that the asymptotic value of function $F_{[\rho]}^2$ at $r_\perp\to\infty$ 
is  $|\Psi_0({\bf R})|^2$, which is spatially inhomogeneous due to the presence of external potential. 
As the result, we obtain equation of motion for ${\bf R}(\beta,t)$ in a variational form,
\begin{equation}
\Gamma[{\bf R}_\beta\times{\bf R}_t]\rho({\bf R})\approx \delta{\cal H}/\delta{\bf R(\beta)}.
\label{variat_R}
\end{equation}

Previously this general structure of vortex filament equation was derived in Ref.\cite{R2001} by using the so called 
vortex line representation for continuously distributed vorticity in anelastic hydrodynamic models,
with subsequent passage to a singular distribution limit. The present derivation naturally takes into account
depletion of the density in vortex core, necessary for correct regularization of Hamiltonian 
${\cal H}\{{\bf R}(\beta)\}$, as well as quantization of the circulation.

The next step in getting Eq.(\ref{LIA}) is simplification of functional ${\cal H}\{{\bf R}(\beta)\}$. 
It is this point where the local induction approximation is applied instead of writing the vortex Hamiltonian as 
a more accurate double o-integral with a regularized  Green's function (see, e.g., Ref.\cite{R2017-2} for more details):
\begin{equation}
{\cal H}\{{\bf R}(\beta)\}\approx {\cal H}_{LIA}=\frac{\Gamma^2\Lambda}{4\pi}\oint\rho({\bf R})|{\bf R}_\beta|d\beta.
\end{equation}
Substitution of this expression into Eq.(\ref{variat_R}) and subsequent resolution with respect to the time derivative
lead us to Eq.(\ref{LIA}).

\begin{figure}
\begin{center}
\epsfig{file=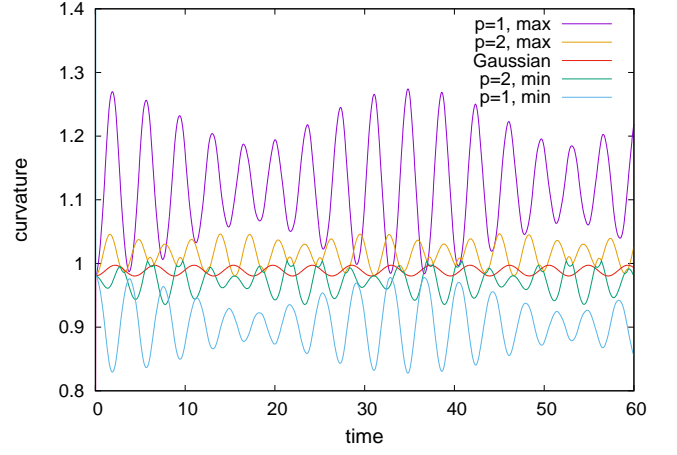, width=88mm}
\end{center}
\caption{Time dependencies of maximal and minimal curvature of initially perfect vortex ring on backgrounds 
$\rho(r,p)=[1-r^2/(1+2p)]^p$, for $p=1$ and for $p=2$, with initial conditions 
$X(\beta,0)= 0.15 + 1.02\cos(\beta), Y(\beta,0)= 1.02\sin(\beta), Z(\beta,0)=0$.
For comparison, curvature of perfect ring on Gaussian background ($p\to\infty$) is presented, 
with the same initial conditions.}
\label{Gauss-parabolic} 
\end{figure}
\begin{figure}
\begin{center}
\epsfig{file=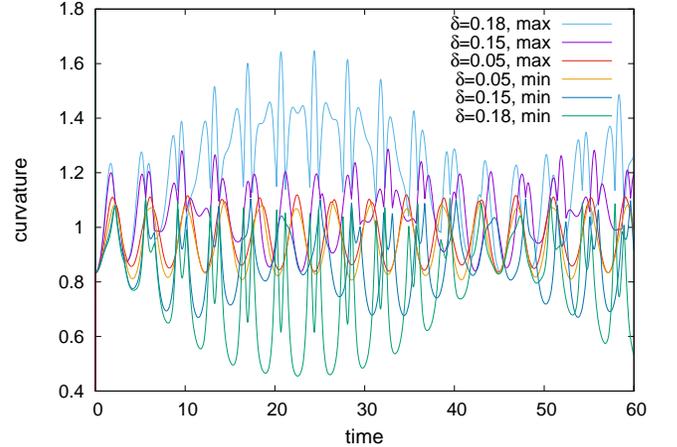, width=88mm}
\end{center}
\caption{Maximal and minimal curvatures of initially perfect vortex ring with 
$X(\beta,0)= 0.3 + 1.2\cos(\beta), Y(\beta,0)=1.2\sin(\beta), Z(\beta,0)=0$, 
for perturbed Gaussian density profiles $\rho(r,\delta)=\exp[-r^2/2 -\delta(r^2-1)^2/4]$.
}
\label{Gauss-distorted} 
\end{figure}

\begin{figure}
\begin{center}
\epsfig{file=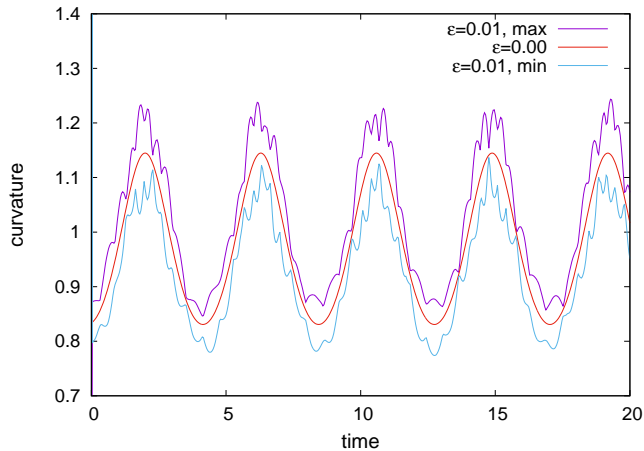, width=88mm}
\end{center}
\caption{Maximal and minimal curvatures of a perturbed vortex ring on Gaussian background with initial conditions
$X(\beta,0)= 0.2 + 1.2\cos(\beta)+ \epsilon\sin(2\beta), Y(\beta,0)=0.1+ (1.2+\epsilon)\sin(\beta), 
Z(\beta,0)=-0.05+\epsilon\cos(2\beta)$, for $\epsilon=0.01$. For comparison, curvature of perfect ring is presented, 
corresponding to $\epsilon=0$.
}
\label{Gauss-ring-perturbed} 
\end{figure}

{\bf Perfect-ring solutions}.
Let us now turn our attention to Gaussian density profile. It is easy to show by a simple geometric consideration 
that the corresponding Eq.(\ref{LIA_Gaussian}) admits a wide class of non-stationary perfect ring configurations. 
Unlike the general $\rho(r)$ case, this class of solutions is far from being exhausted by axisymmetric motion. 
Moreover, if vortex ring of a radius $B$ is directed along a unit (binormal) vector ${\bf b}$ 
and centered at a position ${\bf A}$, then we have the following system of ordinary differential equations for 
two vector functions ${\bf A}(t)$ and ${\bf B}(t)=B{\bf b}$:
\begin{equation}
\dot {\bf A}={\bf B}/|{\bf B}|^2-{\bf B}, \qquad \dot {\bf B}={\bf A}.
\label{AB}
\end{equation}
Apparently, it describes the motion of a Newtonian particle in a central field with potential $W(B)=B^2/2-\ln B$.
The potential has a minimum at $B^2=1$ which is a 2D sphere in the particle's configuration space ${\bf B}$.
Most non-trivial vortex dynamics occurs for solutions with non-zero angular momentum 
${\bf M}=[{\bf B}\times{\bf A}]$ which is an integral of motion. In particular,
a slow regime is possible when the particle moves approximately along the unit sphere. It corresponds to
slow rotation of the vortex ring around ${\bf M}$ direction, accompanied by weak oscillations of radius. 
Of course, qualitatively similar regime is also possible with more general central-symmetric
densities for a slightly distorted ring near equilibrium radius $R_*$ determined by equation 
$\rho(R_*)+R_*\rho'(R_*)=0$. But in Gaussian case the ring keeps perfect shape even for large 
deviations from equilibrium, while bending oscillations are excited on other backgrounds. 
On strongly non-Gaussian densities, especially with sharp boundary, bending oscillations often develop into a
singularity (not shown here).
A difference in behavior of a ``slow'' vortex ring on Gaussian and on some other  backgrounds  
(including parabolic density, corresponding to harmonic trap) is exemplified in Fig.\ref{Gauss-parabolic}, 
based on numerical simulations of Eq.(\ref{LIA}).
In Fig.\ref{Gauss-distorted}, numerical results are presented for initially perfect ring far from equilibrium, 
on weakly non-Gaussian densities. 
A quasi-recurrence is clearly observed in the dynamics, with increasing time period at larger distortions.

The presence of exact integrable reduction (\ref{AB}) distinguishes Eq.(\ref{LIA_Gaussian}) as a very special model
for vortex motion. Therefore some questions arise about Eq.(\ref{LIA_Gaussian}). The first natural question is
if solutions in the form of perfect ring are stable. Numerical simulations of Eq.(\ref{LIA_Gaussian})
demonstrate stability for moderate deviations form the circular shape, as Fig.\ref{Gauss-ring-perturbed} shows. 
More difficult question is if the equation is integrable, or, at least, if there exist some else non-trivial 
finite-dimensional reductions. This question is open at the present moment.

{\bf Conclusions}. Thus, within the local induction approximation, nearly Gaussian equilibrium density 
profiles of Bose-Einstein condensates result in very regular dynamics of quantum vortex rings. Perhaps,
effects of nonlocality in a more accurate vortex Hamiltonian, and even interaction with potential degrees 
of freedom do not destroy this phenomenon. To check this hypothesis, direct numerical simulations of 
the 3D Gross-Pitaevskii equation should be carried out.

\end{document}